\documentclass[prl,twocolumn,superscriptaddress,showpacs,floatfix,longbibliography]{revtex4-1}

\usepackage{epsfig,amsmath,amssymb,latexsym}

\usepackage{subfigure}
\usepackage{xcolor}

\def\kfs122{K$_y$Fe$_2$Se$_2$}
\def\kfsx22{K$_{1-x}$Fe$_2$Se$_2$}

\def\bfa122{BaFe$_2$As$_2$}
\def\fs11{FeSe}

\def\a245{$A_2$Fe$_4$Se$_5$}
\def\neelam{N\'{e}el-AM}
\def\neelaf{N\'{e}el-AF}
\def\neel{N\'{e}el}

\def\rb245{Rb$_2$Fe$_4$Se$_5$}
\def\k245{K$_2$Fe$_4$Se$_5$}
\def\tl245{Tl$_2$Fe$_4$Se$_5$}
\def\cs245{Cs$_2$Fe$_4$Se$_5$}
\def\na245{Na$_2$Fe$_4$Se$_5$}

\def\vecR{\mathbf{R}}

\def\veck{\mathbf{k}}
\def\vecq{\mathbf{q}}

\begin{document}

\title{Spin Triplet Pairing by Suppressing Altermagnetism} \preprint{1}

\author{Xin Ma}
 \affiliation{Center for Correlated Matter and School of Physics, Zhejiang University, Hangzhou 310058, China}

\author{Siqi Wu}
 \affiliation{Department of Physics, The Hong Kong University of Science and Technology, Clear Water Bay, Kowloon, Hong Kong, China}

\author{Zilong Li}
 \affiliation{Center for Correlated Matter and School of Physics, Zhejiang University, Hangzhou 310058, China}

\author{Lunhui Hu}
 \affiliation{Center for Correlated Matter and School of Physics, Zhejiang University, Hangzhou 310058, China}

\author{Jianhui Dai}
 \email[E-mail address: ]{daijh@hznu.edu.cn}
 \affiliation{School of Physics, Hangzhou Normal University, Hangzhou 310036, China}
 \affiliation{Institute for Advanced Study in Physics, Zhejiang University, Hangzhou 310058, China}

%\author{H.-Q. Lin}
% \affiliation{School of Physics, Zhejiang University, Hangzhou 310058, China}
% \affiliation{Institute for Advanced Study in Physics, Zhejiang University, Hangzhou 310058, China}

\author{Chao Cao}
 \email[E-mail address: ]{ccao@zju.edu.cn}
 \affiliation{Center for Correlated Matter and School of Physics, Zhejiang University, Hangzhou 310058, China}
 \affiliation{Institute for Advanced Study in Physics, Zhejiang University, Hangzhou 310058, China}

\date{\today}

\begin{abstract}
The interplay between unconventional superconductivity and altermagnetic order has attracted much attention. In particular, whether spin-triplet superconductivity can be achieved by suppressing altermagnetism remains an open issue. We investigate this issue using a minimal single-orbital Hubbard model on a square lattice with a vacancy superstructure, in which both conventional antiferromagnetic and altermagnetic orders can emerge on an equal footing. We illustrate the existence of a metallic normal phase with altermagnetism even at half-filling due to geometric frustration and Coulomb interaction. Suppressing the altermagnetic long-range order using charge doping can lead to both conventional antiferromagnetic and altermagnetic spin fluctuations. Spin-singlet pairing is always favored when conventional antiferromagnetic spin fluctuations dominate. However, when altermagnetic spin fluctuations dominate, spin-triplet pairing will be induced. Implications of our results for possible material candidates are also briefly discussed.
\end{abstract}

%\pacs{75.25.-j,71.20.-b,75.10.Hk}
\maketitle

%\section{Introduction}
{\it Introduction.} Altermagnetism (AM) has recently been proposed to stand out from conventional antiferromagnetic (AF) orders\cite{PhysRevX.12.040501,PhysRevX.12.031042,PhysRevX.12.011028,PhysRevLett.93.036403,PhysRevB.75.115103,adfm.202409327,Li:2022aa,PhysRevB.109.094425,PhysRevLett.134.106801,Zhang:2025aa,PhysRevLett.134.166701}. Unlike conventional collinear AFs whose Kramers degeneracy is protected by $\lbrace t\vert \mathcal{T}\rbrace$ or $\mathcal{PT}$ symmetry ($t$ is a lattice translation, $\mathcal{P}$/$\mathcal{T}$ denote inversion/time-reversal symmetry), AM features opposite spin sublattices related by a crystallographic symmetry other than inversion or translation\cite{PhysRevX.14.031038,PhysRevX.14.031037,PhysRevB.110.144412,PhysRevB.111.174407,leeb2026collinearpwavemagnetismhidden}, leading to momentum-dependent spin splitting without a net magnetic moment\cite{Krempasky2024-dj,Smejkal2022-wi}. This is highly desirable in spintronics, as even parity ($d$- or $g$-wave) AM Fermi surfaces can give rise to a ``spin-filtering" effect. Thus, AM is established as a scalable, stray-field-free platform for high-density device integration\cite{PhysRevB.110.L140506,52wh-1z5y}.

In addition, the lifted Kramers degeneracy renders the interplay between AM and superconductivity (SC) intriguing\cite{Mazin:2025aa}. In principle, there are three distinct scenarios for the AM/SC interplay: (i) The SC phase is microscopically separated from the AM phase, i.e., the SC/AM system forms junctions or interfaces.  In such cases, studies have investigated the proximity effect between altermagnets and conventional superconductors, where it is reported that finite-momentum pairing without external fields \cite{zhang2024finite}, pseudo-Ising SC induced by odd-symmetry AM phases\cite{cm1l-1rsh}, and topological superconducting phases driven by the altermagnetic proximity effect \cite{rhmg-j1fv,kqy8-myz1} may occur. In addition, subgap states may be present\cite{fz2b-5sp7}, and triplet SC via Floquet engineering\cite{lkf9-jgv6} and extensive studies have been conducted on the Andreev reflection and Josephson junction effects\cite{PhysRevLett.131.076003,PhysRevB.108.L060508,Kazmin_2025}. (ii) SC microscopically coexists with AM. For even-parity AM phases, since the singlet pairing between $|\veck\uparrow\rangle$ and $|-\veck\downarrow\rangle$ states is forbidden, triplet pairing is naturally expected\cite{52wh-1z5y,PhysRevB.108.224421,PhysRevB.109.134515,PhysRevB.110.L220503,g4dl-1ff2,PhysRevB.111.054501,parshukov2025exoticsuperconductingstatesaltermagnets,mineev2026toroidaltermagneticnoncentrosymmetricordering,wzch-zzwg,yokoyama2025relationshipnonunitarymixedparity}, though the Fulde-Ferrell-Larkin-Ovchinnikov (FFLO) state is proposed to be energetically more favorable\cite{cv8s-tk4c,hu2025unconventionalsuperconductivityaltermagneticmetal,Hu_2025,3k12-2467,liu2026altermagnetismdrivenfflosuperconductivityfinitefilling,hu2026finitemomentumsuperconductivitysinglettripletmixing,Mukasa_2025}. (iii) SC arises from a paramagnetic state upon suppressing the long-range AM order. Since SC coexisting with magnetic order is quite rare, the last scenario is more likely to be realized. However, it remains significantly less explored. 

While the lifted Kramers degeneracy in the first two scenarios naturally disfavors conventional spin-singlet pairing, the same argument does not directly apply to the last one. This is already illustrated by previous studies, where the SC phase remains dominated by $s$-wave or $d$-wave singlet pairings\cite{PhysRevB.110.205120}, and short-range fluctuations favor intra-cell singlet pairings \cite{dlpb-gfct}. These studies employ $t-J$ models, in which the forms of spin fluctuations are predetermined by the magnetic exchange couplings $J$. In unconventional SCs, the pairing symmetry of SC is dictated by the dynamic spin fluctuations in the paramagnetic state\cite{RevModPhys.63.239,RevModPhys.84.1383,RevModPhys.78.17}. For example, spin-singlet pairing is closely related to the AF spin fluctuations in cuprates\cite{SCALAPINO1995329,RevModPhys.84.1383}, iron-pnictides\cite{Hirschfeld_2011,RevModPhys.84.1383}, and most recently, nickelates \cite{PhysRevLett.131.126001,Xia:2025aa,PhysRevB.109.235126}. In contrast, if the system is dominated by ferromagnetic (FM) spin fluctuations, spin-triplet SC may occur, as proposed in UTe$_2$ \cite{science.aav8645}. Conventionally, AF and FM orders are characterized by ordering wave-vectors $\mathbf{Q}\neq \Gamma$ and $\mathbf{Q}=\Gamma$, and net moments $\mathbf{M}=0$ and $\mathbf{M}\neq 0$, respectively. In contrast, AM has an ordering wave-vector $\mathbf{Q}=\Gamma$ and a net moment $\mathbf{M}=0$. As a result, it is particularly compelling to investigate whether spin-triplet SC can emerge from a paramagnetic normal state when the AM long-range order is suppressed \cite{Mazin:2025aa}. However, once the AM long-range order is suppressed, a mixture of AM, AF, and FM spin fluctuations may emerge, and the dominant channel may not be AM. Therefore, to systematically determine the dominant fluctuation and the pairing symmetry, it is critical to employ a microscopic model that treats each magnetic order on an equal footing, without assuming additional local spins or order parameters.

In this {\it Letter}, we perform mean-field calculations and a random-phase approximation (RPA) analysis of a single-orbital Hubbard-like model on a $\sqrt{5}\times\sqrt{5}$ vacancy-ordered square lattice with $I4/m$ symmetry, which was originally motivated by a class of iron chalcogenide superconductors\cite{Fang_2011}. We show that conventional AF orders and a $d$-wave AM order can spontaneously emerge in the large-$U$ regime, without a pre-assumed order parameter. Remarkably, the AM order addressed here is {\it intrinsic} in that the ordering vector is $\mathbf{Q}=\Gamma$ even for the magnetic-site-only lattice, and therefore cannot be represented by downfolded conventional AF orders. Meanwhile, in the intermediate interaction range, a metallic AM phase may be present even at half-filling. Upon suppressing the AM long-range order using charge doping, both conventional AF and AM fluctuations emerge. When conventional AF fluctuations dominate, spin-singlet pairing is always favored. However, when AM fluctuations dominate, spin-triplet pairing can be induced for sufficiently strong interactions.

\begin{figure*}[htp]
 \includegraphics[width=16cm]{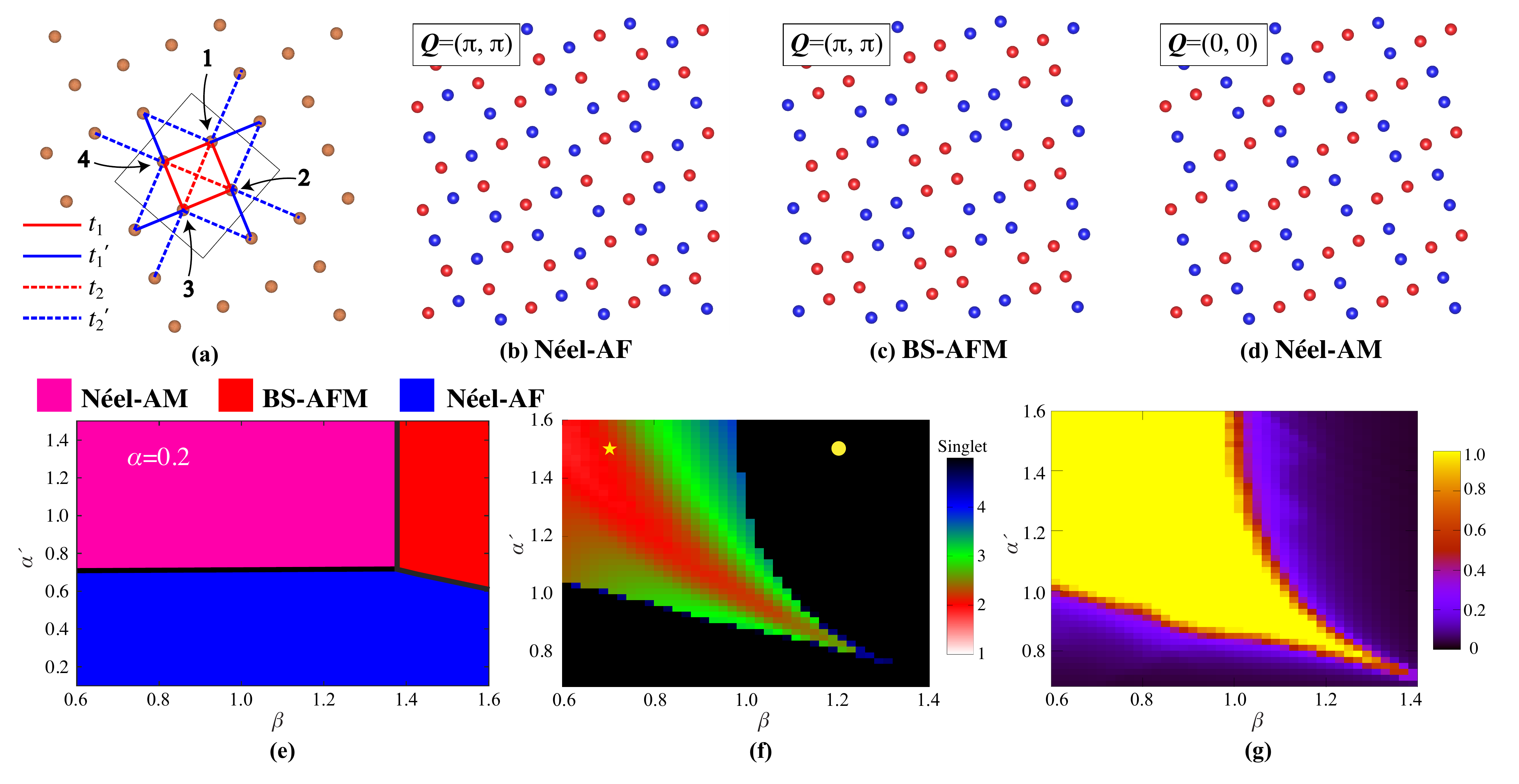}
 \caption{(a) $\sqrt{5}\times\sqrt{5}$ vacancy-ordered square lattice. The red/blue lines indicate intra-block/inter-block hoppings; and the solid/dashed lines indicate nearest-neighboring/next-nearest-neighboring hoppings. (b-d) show the \neelaf\ order, BS-AFM order, and \neelam\ order, respectively. Up/down moments are represented by red/blue colors. (e) Magnetic phase diagram at $U=8$ at half-filling. (f) The minimum interaction $U_t$ when the triplet becomes dominant for the $0.02e$/site-doped system. The black area indicates dominant singlet pairing only. The positions of the two cases discussed in detail are marked with the star and circle. (g) Dominant magnetic spin fluctuation at $U = 0.99U_c$, indicated by the ratio $\gamma$ between the strengths of AM-type and dominant spin fluctuations. $\gamma=1$  means the system is dominated by AM-type spin fluctuations. $\alpha=0.2$ for (e-g).\label{fig:phase_diagram}}
\end{figure*}

{\it Microscopic Model.} We consider a two-dimensional square-lattice Hubbard model $H=H_0+H_{\mathrm{int}}$ with a perfect $\sqrt{5}\times\sqrt{5}$ vacancy superstructure, as shown in FIG. \ref{fig:phase_diagram}a. For simplicity, we consider only hoppings up to next-nearest-neighboring (n.n.n.) sites and only one orbital per site. Due to the existence of the vacancy superstructure, the lattice can be viewed as a composite square lattice with a square block of four internal sites at each lattice point. Therefore, the non-interacting Hamiltonian can be written as:
\begin{eqnarray}
H_0 &=& \sum_{\vecR,\sigma} \left( t_1 \sum_{\langle i,j\rangle} c_{\vecR i\sigma}^{\dagger} c_{\vecR j\sigma} + t_2 \sum_{\langle\langle i,j\rangle\rangle} c_{\vecR i\sigma}^{\dagger} c_{\vecR j\sigma} \right) \nonumber \\
    &+& \sum_{\langle\vecR,\vecR'\rangle,\sigma} \left( t'_1 \sum_{\langle i,j\rangle'} c_{\vecR i\sigma}^{\dagger} c_{\vecR' j\sigma} + t'_2 \sum_{\langle\langle i,j\rangle\rangle'} c_{\vecR i\sigma}^{\dagger} c_{\vecR' j\sigma} \right) \nonumber \\
    &-& \mu \sum_{\vecR i\sigma} \hat{n}_{\vecR i\sigma} 
 \label{eq:ham0}
\end{eqnarray}
where $\vecR$ and $\vecR'$ denote the composite lattice points of square blocks; $\langle\vecR,\vecR'\rangle$ are nearest-neighboring (n.n.) blocks; $1\leqslant i,j \leqslant4$ are the internal site indices (FIG. \ref{fig:phase_diagram}a); $\langle i,j\rangle$ and $\langle i,j\rangle'$ indicate intra-block and inter-block n.n. sites, respectively; $\langle\langle i,j\rangle\rangle$ and $\langle\langle i,j\rangle\rangle'$ indicate intra-block and inter-block n.n.n. sites, respectively. The $c_{\vec{R},i\sigma}^{\dagger}$ ($c_{\vec{R},i\sigma}$) are the creation (annihilation) operators for spin $\sigma$ at site $i$ of unit cell $\vec{R}$. In addition, we introduce the on-site interactions:
\begin{eqnarray}
H_{\mathrm{int}}=\sum_{\vecR,i}U\hat{n}_{\vecR i\uparrow}\hat{n}_{\vecR i\downarrow}
 \label{eq:hamint}
\end{eqnarray}
It is clear that our microscopic model does not explicitly introduce an altermagnetic order parameter. As a result, both conventional AF and AM fluctuations can be studied on an equal footing. Without loss of generality, we shall fix $t_1=1.0$ in the following discussion.

Using the standard mean-field decomposition\cite{PhysRevB.35.3359}, the half-filling ground state phase diagram of this model was investigated in Ref. \cite{PhysRevB.83.180413}.  In general, the model shows vacancy-enhanced metal-insulator transitions and magnetic orderings by increasing the Coulomb interaction $U$. The dominant magnetic configurations, such as the conventional AF (FIG. \ref{fig:phase_diagram}b) and the block-spin AF (BS-AFM) (FIG. \ref{fig:phase_diagram}c) states, depend delicately on the intra-block and inter-block hopping frustrations\cite{PhysRevB.83.180413}. Remarkably, the AM configuration (FIG. \ref{fig:phase_diagram}d) can also emerge in this model on an equal footing, as we shall illustrate below. Here, we present the typical magnetic phase diagram at half-filling with $U=8$ (FIG. \ref{fig:phase_diagram}e). In the large-$U$ limit, the hoppings lead to AF exchanges between n.n. and n.n.n. [$J_i\approx 4t^2_i/U$, $J'_i\approx 4t'^2_i/U$, ($i=1,2$)], resulting in an extended $J_1-J_2$ model\cite{PhysRevB.83.180413,PhysRevB.84.094451}, and the ratio of $t_2/t_1$ implies the strength of the intra-block frustration. When the intra-block frustration is small, e.g., $\alpha=t_2/t_1=0.2$, our mean-field calculation identifies three major magnetic phases at half-filling: i) if $\alpha'=t'_2/t'_1$ is also small, both intra- and inter-block n.n. exchange interactions dominate, and the \neelaf\ phase is favored; ii) if $\alpha'$ is large and $\beta=t'_1/t_1$ is small, $J_1$ and $J'_2$ dominate intra- and inter-block interactions, and the \neelam\ phase is favored; iii) if $\alpha'$ is large but $\beta=t'_1/t_1$ is large, the inter-block interactions $J'_1$ and $J'_2$ dominate, and the BS-AFM phase is favored. We stress here that these phase diagrams are determined at $U=8$; thus, they do not necessarily reflect the properties of the leading spin fluctuations in the paramagnetic state. We also note that the non-interacting system ($U=0$) is always metallic in the parameter phase space at half-filling\cite{PhysRevB.83.180413}. If the intra-block frustration $\alpha$ is reduced/increased, the space of the \neelam\ phase will expand/shrink, but these major phases can always be identified for $\alpha\leqslant 0.6$ \cite{note_SM}. In the following discussions, we focus on the small frustration case ($\alpha=0.2$).

%In order to investigate the SC properties, we suppress the magnetic long-range order by doping the system with 0.02e/site.  
{\it Superconducting Phase Diagram.} In order to investigate the SC properties, we consider $\beta\in [0.6, 1.4]$ and $\alpha'\in [0.68, 1.6]$, which cover the \neelam\ region obtained from the mean-field results at $U=8$. The suppression of magnetic long-range order is simulated by considering charge doping. To achieve this, the critical interaction strengths for the magnetic transition both at half-filling ($U_{c1}$) and with charge doping ($U_{c2}$) are determined within RPA, beyond which the system is regarded as being in the magnetically ordered state. For a small deviation from half-filling, $U_{c1}<U_{c2}$ is usually satisfied; thus, within $U_{c1}\leqslant U <U_{c2}$ the long-range magnetic order at half-filling is suppressed by the charge doping. The pairing strengths and gap functions within $[0.9U_{c1}, U_{c2})$ are then calculated. In particular, the bare electron susceptibility tensor for the non-interacting system in the particle-hole channel is given by:
\begin{eqnarray}
\left[\chi_0(\omega, \vecq)\right]^{pq}_{st}&=&-\frac{1}{N_{\veck}} \sum_{\veck\mu\nu}\frac{f(\epsilon_{\nu\veck+\vecq})-f(\epsilon_{\mu\veck})}{\omega+\epsilon_{\nu\veck+\vecq}-\epsilon_{\mu\veck}+i0^+} \nonumber \\
 &\times&\langle s |\mu\veck\rangle \langle \mu\veck | p\rangle\langle q|\nu\veck+\vecq\rangle\langle \nu\veck+\vecq | t\rangle
 \label{eq:chi0}
\end{eqnarray}
where $|\mu\veck\rangle$ and $\epsilon_{\mu\veck}$ are the $\mu$-th Bloch state at $\veck$ and the corresponding eigenenergy; $f$ is the Fermi-Dirac function; $s, p, q, t$ are orbital (site) indices; and the summation takes place in the first Brillouin zone (BZ) sampled by a $256\times256$ $\veck$-mesh.

Within the random-phase approximation (RPA), the general susceptibility tensor $\chi^{ch}$ is then obtained by solving $\chi^{ch}=\chi_0 + \chi^{ch} U^{ch}\chi_0$, where $ch$ can be $S$ for the spin channel or $C$ for the charge channel, and $U^{ch}$ is the corresponding interaction tensor\footnote{For the single-orbital on-site case, $\left[U^S\right]_{ii}^{ii}=-\left[U^C\right]_{ii}^{ii}=U$, and all other interaction tensors are zero.}. The magnetic transition occurs at $U_c$, where the largest eigenvalue of $[U^S\chi_0]$ is 1. For $U<U_c$, in the weak coupling limit, the spin-singlet ($V^s$)/triplet ($V^t$) channels of the pairing interactions can be calculated using:
\begin{eqnarray}
 \left[V^s(\veck', \veck)\right]^{ii}_{jj}&=& \left[U^S\right]^{ii}_{jj}+\frac{1}{4}\lbrace [3U^S \chi^S(\veck'-\veck)U^S \nonumber \\
  &-& U^C \chi^C(\veck'-\veck)U^C]_{jj}^{ii}+(\veck'\leftrightarrow\veck)\rbrace
 \label{eq:vsc_s}
\end{eqnarray}
\begin{eqnarray}
 \left[V^t(\veck', \veck)\right]^{ii}_{jj}&=&-\frac{1}{4}\lbrace[U^S \chi^S(\veck'-\veck)U^S \nonumber \\
  &+& U^C \chi^C(\veck'-\veck)U^C]_{jj}^{ii}-(\veck'\leftrightarrow\veck)\rbrace
 \label{eq:vsc_t}
\end{eqnarray}
where $i$ and $j$ are the orbital (site) indices without spin. All other terms are zero since we consider only on-site interactions. By solving the linearized gap equation:
\begin{eqnarray}
\lambda \Delta(\veck')=-\frac{1}{V_{\mathrm{BZ}}}\int_{\mathrm{FS}}\frac{d^2 k_{\parallel}}{|v_{\veck}^{\perp}|}V(\veck', \veck)\Delta(\veck)
 \label{eq:gapeq}
\end{eqnarray}
the pairing strengths $\lambda$ of all channels are obtained. For the model of interest, there are two singlet channels ($A_g$ and $B_g$) and one two-dimensional triplet channel ($E_u$). 

In FIG. \ref{fig:phase_diagram}f, we show the minimum interaction strength $U_t$ when the triplet channel becomes dominant for the $0.02e$/site-doped system. $U_t$ is defined as the interaction when the triplet/singlet pairing strengths become equal, i.e., $\mathrm{max}(\lambda^{A_g}, \lambda^{B_g})=\lambda^{E_u}$ at $U_t$. The black area indicates that $\lambda^{E_u}<\mathrm{max}(\lambda^{A_g}, \lambda^{B_g})$ for $U\leqslant 0.99U_c$, namely dominant singlet pairing only. Inside the \neelam\ region, a triplet superconducting fan is evident. However, the area of the triplet pairing phase is significantly smaller than that of the \neelam\ phase, and singlet pairing still dominates close to the phase boundaries. To further investigate, we choose two typical cases for closer examination: 1) deep in the \neelam\ region ($\alpha=0.2$, $\beta=0.7$, $\alpha'=1.5$) and 2) close to the BS-AFM phase boundary ($\alpha=0.2$, $\beta=1.2$, $\alpha'=1.5$). 

%We calculated the difference $\Delta \lambda=\lambda^{t}_{\mathrm{m}} - \lambda^{s}_{\mathrm{m}}$ between the leading triplet eigenvalue $\lambda^{t}_{\mathrm{m}}$ and the leading singlet eigenvalue $\lambda^{s}_{\mathrm{m}}$ at $U=0.99U_c$ for $\beta\in [0.6, 1.4]$, $\alpha'\in [0.68, 1.6]$ in the small frustration region for 0.02e/site doped case (FIG. \ref{fig:phase_diagram}h), covering the \neelam\ phase region in the large-$U$ limit. A positive $\Delta\lambda$ indicate dominant triplet pairing instability. Inside the \neelam\ region, a triplet superconducting fan is evident. However, the area of the triplet pairing phase is significantly smaller than the \neelam\ phase, and singlet pairing still dominates close to the phase boundaries. Therefore, we chose 2 typical cases for closer examination: 1) deep in the \neelam\ region ($\alpha=0.2$, $\beta=0.7$, $\alpha'=1.5$) and 2) close to the BS-AFM phase boundary ($\alpha=0.2$, $\beta=1.2$, $\alpha'=1.5$). 

%, and 3) close to the \neelaf\ phase boundary ($\alpha=0.2$, $\beta=0.8$, $\alpha'=0.75$).

\begin{figure*}[htp]
 \includegraphics[width=16cm]{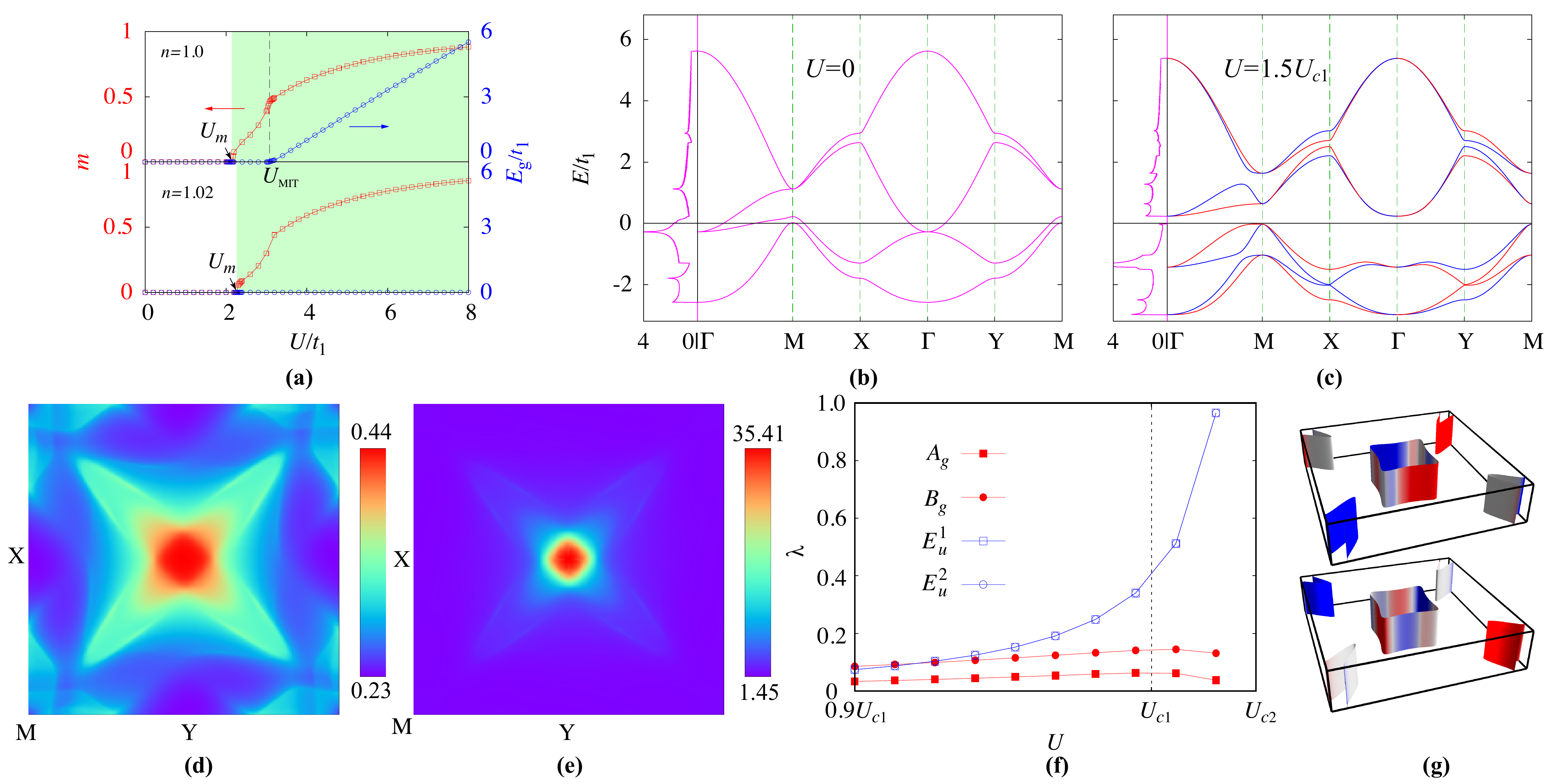}
 \caption{(a) Magnetic moment per site $m$ and energy gap $E_g$ under different $U$ for (upper-panel) half-filling ($n=1.0$) and $0.02e$ doped ($n=1.02$) systems. The shaded area corresponds to the \neelam\ phase. The vertical dashed line indicates the metal-insulator transition. (b-c) Band structure and DOS of (b) paramagnetic phase ($U$=0) and (c) \neelam\ phase ($U=1.5U_{c1}$) at half-filling. In (c), red/blue color indicates up/down spins, respectively. (d) Largest eigenvalue of the bare electron susceptibility $\chi_0$. (e) The RPA susceptibility $\chi^{S,\mathrm{N}}$ at $U=0.95U_{c2}$. (f) Leading eigenvalues of the linearized gap equations. (g) The gap function $\Delta(\veck)$ of the two degenerate $E_u$ representations ($U=2.14$). Red/blue denotes positive/negative values.\label{fig:case1}}
\end{figure*}

{\it Deep in the \neelam\ Region.} We first identify the phase evolution close to half-filling with increasing $U$ using mean-field calculations. The non-interacting system ($U$=0) is metallic and spin-degenerate (FIG. \ref{fig:case1}a-b). At half-filling, a magnetic phase transition and a subsequent metal-insulator transition (MIT) occur at $U_m\approx2.18$ and $U_{\mathrm{MI}}\approx3.08$ (FIG. \ref{fig:case1}a), respectively. We note that our mean-field calculations do not enforce any spatial symmetry. Nevertheless, the converged ground state in this case always remains symmetric under $\lbrace C_4 | \mathcal{T}\rbrace$ throughout $U\in [0, 8]$. After the magnetic transition at $U_m$, the local moment per site $m$ quickly increases and reaches approximately half of its maximum value before the MIT. Between $U_{\mathrm{MI}}$ and $U=8$, the gap size $E_g$ increases linearly from 0 to $5.52$, and $m$ gradually approaches saturation. The spin-split band structure can always be observed for $U>U_m$, confirming the AM ground state (FIG. \ref{fig:case1}c). Under $0.02e$/site electron doping, the MIT is completely suppressed, and $U_m$ is also slightly enhanced to $2.30$. Nevertheless, the magnetic ground state is still the \neelam\ phase for $U>U_m$. Therefore, one would expect \neelam -type spin fluctuations to be dominant once the long-range order is suppressed.

%($\mathrm{max}\lbrace\lambda_0\rbrace=0.4426$) 
Such expectation is confirmed by our RPA calculations. Under RPA, the magnetic transition occurs at $U_{c1}=2.18$ at half-filling. Electron doping of $0.02e$/site slightly enhances the transition to $U_{c2}=2.26$. FIG. \ref{fig:case1}d shows the largest eigenvalues of the static $\chi_0(\omega=0)$ in the first BZ. A dominant peak at $\Gamma$ is apparent, indicating that the most susceptible channel is either ferromagnetic (FM) or AM. The eigenvector of the leading eigenvalue corresponds to $\hat{P}=\frac{1}{2}\sum_i (-1)^i c^{\dagger}_i c_i=\frac{1}{2}\left(-c^{\dagger}_1 c_1 + c^{\dagger}_2c_2 - c^{\dagger}_3c_3 + c^{\dagger}_4c_4\right)$. We note that in the spin channel, it represents the intra-block \neel\ order at $\Gamma$, corresponding to the \neelam -type spin fluctuations. When on-site interactions are included, the charge fluctuations are suppressed in RPA, and the AM-type spin fluctuations dominate.

To verify this, we directly calculate the intra-block \neel /FM-channel spin susceptibility using:
\begin{eqnarray}
\chi^{S,\nu}(\vecq)=\sum_{spqt} \left[ O^{\nu} \right]_{st} \left[ O^{\nu} \right]_{pq} \left[\chi^{S}(\vecq)\right]^{pq}_{st}
\label{eq:chich}
\end{eqnarray}
where $\nu$ can be $\mathrm{N}$, $\mathrm{F}$, and $\mathrm{L}$ for \neel, FM, and stripe-like channels, with $O^{\mathrm{N}}_{pq}=(-1)^p\delta_{pq}$, $O^{\mathrm{F}}_{pq}=\delta_{pq}$ and $O^{\mathrm{L}}_{pq}=\xi_p\delta{pq}$ ($\xi=-1$ for $p=1,2$; $\xi=1$ for $p=3,4$), respectively. The $\chi^{S,\mathrm{N}}$ exhibits a strong peak at $\Gamma$ (FIG. \ref{fig:case1}e), which is an order of magnitude larger than the $\chi^{S,\mathrm{F}}$ peak\cite{note_SM}, indicating that the dominant spin fluctuation in the system is AM-like. We also note that the peak of $\chi^{S,\mathrm{F}}$ is not located at $\Gamma$ but close to M $(\pi, \pi)$; therefore, it indicates BS-AFM-type fluctuations instead of FM fluctuations.

 FIG. \ref{fig:case1}f shows the leading eigenvalues for different pairing symmetries. Since we are interested in the SC after suppressing the long-range-ordered magnetic states, we solve the gap equation from slightly below $U_{c1}$ to $U_{c2}$. At $0.9U_{c1}$, the leading channel is spin-singlet $B_g$. However, the $E_u$ channel is very close and diverges much faster. Between $U_{c1}$ and $U_{c2}$, the triplet $E_u$ pairing is dominant. It is worth noting that the $E_u$ representation is two-dimensional, with nodal lines on the two Fermi pockets (FIG. \ref{fig:case1}g). The nodal lines of $E^1_{u}$ and $E^2_{u}$ are located at different positions; therefore, these two gap functions may mix to form a fully gapped $p_x+ip_y$ superconducting state to further reduce the total energy. 
 
\begin{figure*}[htp]
 \includegraphics[width=16cm]{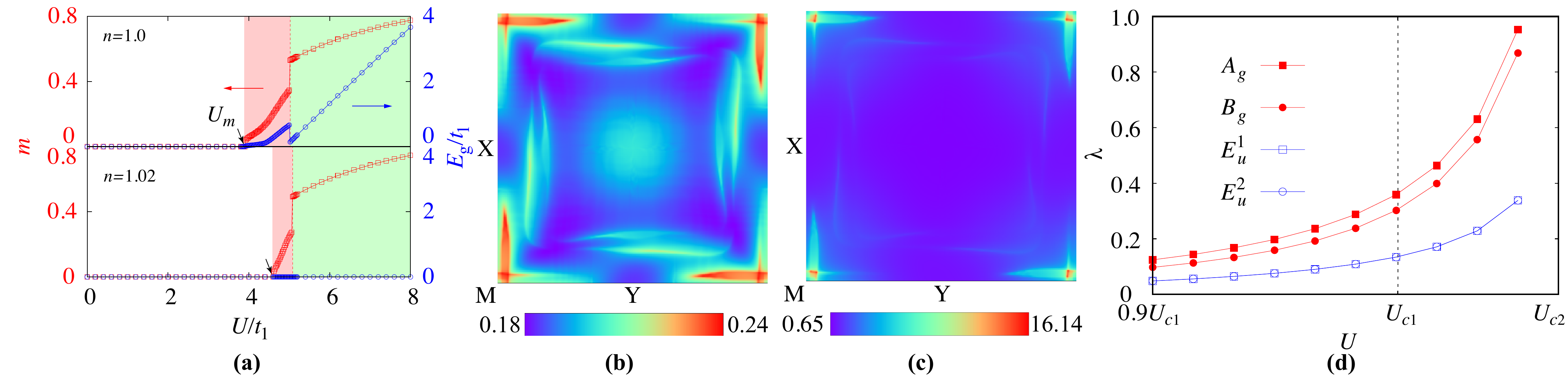}
 \caption{(a) Magnetic moment per site $m$, energy gap $E_g$ of half-filling and $0.02e$ doped system close to the phase boundary of the BS-AFM phase. The green and red shaded areas correspond to the \neelam\ and BS-AFM phases. (b) The largest eigenvalue of the bare electron susceptibility $\chi_0$. (c) The RPA susceptibility $\chi^{S,\mathrm{F}}$ at $U=0.95U_{c2}$ for the $0.02e$ doped system. (d) Leading eigenvalues of the linearized gap equations.\label{fig:case23}}
\end{figure*}

{\it Close to the Phase Boundary.} We now examine the case when the system is close to the boundary of the BS-AFM phase. In the mean-field calculations, we found that although the magnetic ground state is \neelam\ under large $U$ at half-filling, the ground state is BS-AFM in the intermediate $U$ range ($U_m\leqslant U < U'_m$) (FIG. \ref{fig:case23}a). At half-filling, the formation of magnetic long-range order immediately opens the band gap, namely $U_{\mathrm{MIT}}=U_m$. At $U'_m$, another magnetic phase transition occurs, and the system enters the \neelam\ phase. The second magnetic phase transition is signaled by a sudden increase in $m$ and a drop in the energy gap $E_g$, which then increases linearly to $3.67$ at $U=8.0$. With $0.02e$ electron doping, the MIT is again completely suppressed, and both $U_m$ and $U'_m$ are increased (FIG. \ref{fig:case23}a). Nevertheless, the size of the intermediate BS-AFM phase remains finite. Therefore, suppressing the \neelam\ long-range order in this case does not necessarily lead to AM-type spin fluctuations.

The RPA calculation also confirms that the \neelam -type spin fluctuations do not dominate in this case. Specifically, the magnetic transitions are identified at $U_{c1}=3.89$ for half-filling and $U_{c2}=4.14$ for the $0.02e$/site-doped system. Approaching this instability, the largest eigenvalue of $\chi_0$ exhibits a peak around M $(\pi, \pi)$ instead of at $\Gamma$ (FIG. \ref{fig:case23}b). At M, the eigenvector of the largest eigenvalue corresponds to $\hat{Q}=\frac{1}{2}\sum_i c^{\dagger}_i c_i=\frac{1}{2}\left(c^{\dagger}_1c_1 + c^{\dagger}_2c_2 + c^{\dagger}_3c_3 + c^{\dagger}_4c_4\right)$. This represents the intra-block FM order in the spin channel; thus, the peak at $M$ corresponds to BS-AFM-type spin fluctuations. Therefore, when on-site interactions are included, BS-AFM-type spin fluctuations dominate, as shown by the peak of $\chi^{S,\mathrm{F}}$ around M (FIG. \ref{fig:case23}c)\cite{note_SM}. Therefore, the system exhibits spin-singlet $A_g$ pairing instead (FIG. \ref{fig:case23}d).

%At M, there are also 4 degenerate eigenvectors for the largest eigenvalue, which are $Q^{\pm}=\frac{1}{2\sqrt2}\sum_i \left[c^{\dagger}_{i\uparrow} c_{i\uparrow} \pm c^{\dagger}_{i\downarrow} c_{i\downarrow}\right]$ and $R^{\pm}=\frac{1}{2\sqrt2}\sum_i \left[c^{\dagger}_{i\uparrow} c_{i\downarrow} \pm c^{\dagger}_{i\downarrow} c_{i\uparrow}\right]$. In contrast to the previous case, $Q^-$ is proportional to intra-block FM order parameter, and is significantly enhanced when on-site interactions are included. The dominant fluctuation in this case is BS-AFM type, as shown by the peak of $\chi^{\mathrm{F}}$ around M (right panel of FIG. \ref{fig:case23}b). Therefore, the system exhibits spin-singlet $A_g$ pairing instead (FIG. \ref{fig:case23}c).

{\it Discussion and Conclusion.} Our results above show that if the \neelam\ long-range order is suppressed, both AM-type spin fluctuations and conventional AF spin fluctuations exist in the system. If conventional AF fluctuations dominate, spin-singlet pairing still dominates. This also applies to the system near the \neelaf -ordered phase boundary. In fact, both the mean-field calculation and RPA analysis show that the dominant fluctuations in that case are not \neelam -type either, eventually leading to singlet SC\cite{note_SM}. Thus, spin-triplet SC can only be realized if the AM spin fluctuations dominate.

To further demonstrate the correlation between the \neelam-type spin fluctuations and spin-triplet pairing, we also plotted $$\gamma=\chi^{S,\mathrm{N}}(\Gamma)/\max \lbrace\chi^{S,\mathrm{N}}(\vecq),\,\chi^{S,\mathrm{F}}(\vecq),\,\chi^{S,\mathrm{L}}(\vecq)\rbrace$$, where the numerator and denominator represent the strength of the \neelam-type spin fluctuations and the maximum strength among all competing spin fluctuations. Therefore, $\gamma=1$ indicates AM-type spin fluctuations are dominant. In FIG. \ref{fig:phase_diagram}g, the phase space of dominating AM-type spin fluctuations coincides with the spin-triplet pairing phase space, suggesting that leading AM spin fluctuations result in spin-triplet pairing.

We note that the vacancy-ordered \mbox{$\sqrt{5}\times\sqrt{5}$} lattice provides a promising platform that can be realized in ternary layered \mbox{$A_2$Fe$_4$Se$_5$} compounds \mbox{\cite{Fang_2011,Cai_2013,Bao_2015,PhysRevB.100.094108}}. Earlier studies have identified an insulating BS-AFM ground state for this material at ambient pressure \mbox{\cite{Bao_2011,PhysRevLett.107.056401,PhysRevB.83.233205}}, and a re-emergent SC phase at high pressure \mbox{\cite{Sun:2012aa}}. Our present study reveals that an AM order can intrinsically emerge in this or other \mbox{$\sqrt{5}\times\sqrt{5}$} lattice materials. Upon suppressing the AM order via electron doping or physical pressure, spin-triplet superconducting pairing could be realized in the vicinity of the AM quantum critical point, driven by the strong AM fluctuations. In addition, similar structure can be identified in other crystals with I4/m symmetry, including Ba$_2$CuOsO$_6$, Nb$_5$(CuSi)$_4$, {\it et al.} Our model can also be realized in a decorated square lattice\cite{PhysRevB.110.205120,PhysRevLett.134.166701}, or a square lattice with internal degrees of freedom. Additionally, when considering interband effects\cite{PhysRevLett.132.036001}, although tuning $t_2$ leads to a simultaneous variation in $U_t$, the triplet superconducting phase remains robust when the frustration is not too large\cite{note_SM}. We also note the altermagnetism and the resulting spin-triplet pairing rely on the underlying spatial symmetry and hopping parameters. A strong disorder effect may disrupt the crystal symmetries, leading to the emergence of in-gap bound states \mbox{\cite{PhysRevB.111.174436,f6nc-vsnx}}, or even driving a metal-insulator transition\mbox{\cite{sw4z-f5fl}}. 

In conclusion, we performed mean-field and RPA calculations of a minimal microscopic model which treats conventional AF and altermagnetism on an equal footing. We show that suppressing the altermagnetic long-range order can lead to both conventional AF and altermagnetic spin fluctuations. If the conventional AF spin fluctuations dominate, spin-singlet pairing is always favored. In contrast, spin-triplet pairing can be realized if the altermagnetic spin fluctuations dominate and the interaction is sufficiently strong.  This opens a possible route to realize triplet pairing SC in compounds with altermagnetism.

\begin{acknowledgments}
We would like to thank Hua Chen, Chenchao Xu, Ming Shi, Yang Liu and Yu Song for the stimulating discussions. This work has been supported by the National Key R\&D Program of China (Nos. 2024YFA1408303 \& 2022YFA1402202) and the National Natural Science Foundation of China (Nos. 12274364 \& 12274109). The calculations were performed on the Quantum Many-Body Computing Cluster at Zhejiang University and High Performance Computing Center of Hangzhou Normal University.
\end{acknowledgments}

 \bibliography{model}

\end{document}